\let\includefigures=\iftrue
%
\let\useblackboard=\iftrue
%
%
\newfam\black
\input harvmac.tex
\input epsf
\includefigures
\message{If you do not have epsf.tex (to include figures),}
\message{change the option at the top of the tex file.}
\def\figin{\epsfcheck\figin}\def\figins{\epsfcheck\figins}
\def\epsfcheck{\ifx\epsfbox\UnDeFiNeD
\message{(NO epsf.tex, FIGURES WILL BE IGNORED)}
\gdef\figin##1{\vskip2in}\gdef\figins##1{\hskip.5in}
\else\message{(FIGURES WILL BE INCLUDED)}%
\gdef\figin##1{##1}\gdef\figins##1{##1}\fi}
\def\DefWarn#1{}
\def\figinsert{\goodbreak\midinsert}
\def\ifig#1#2#3{\DefWarn#1\xdef#1{fig.~\the\figno}
\writedef{#1\leftbracket fig.\noexpand~\the\figno}%
\figinsert\figin{\centerline{#3}}\medskip\centerline{\vbox{\baselineskip12pt
\advance\hsize by -1truein\noindent\footnotefont{\bf Fig.~\the\figno:} #2}}
\bigskip\endinsert\global\advance\figno by1}
\else
\def\ifig#1#2#3{\xdef#1{fig.~\the\figno}
\writedef{#1\leftbracket fig.\noexpand~\the\figno}%
\global\advance\figno by1}
\fi
\useblackboard
\message{If you do not have msbm (blackboard bold) fonts,}
\message{change the option at the top of the tex file.}
\font\blackboard=msbm10 scaled \magstep1
\font\blackboards=msbm7
\font\blackboardss=msbm5
\textfont\black=\blackboard
\scriptfont\black=\blackboards
\scriptscriptfont\black=\blackboardss
\def\Bbb#1{{\fam\black\relax#1}}
\else
\def\Bbb{\bf}
\fi
%
\def\yboxit#1#2{\vbox{\hrule height #1 \hbox{\vrule width #1
\vbox{#2}\vrule width #1 }\hrule height #1 }}
\def\fillbox#1{\hbox to #1{\vbox to #1{\vfil}\hfil}}
\def\ybox{{\lower 1.3pt \yboxit{0.4pt}{\fillbox{8pt}}\hskip-0.2pt}}

\def\comments#1{}

\def\QC{\Bbb{C}}

\def\QP{\Bbb{P}}
\def\QR{\Bbb{R}}

\def\QZ{\Bbb{Z}}

\def\CA{{\cal A}}

\def\CF{{\cal F}}

\def\II{\relax{I\kern-.10em I}}

\def\IZ{\relax\ifmmode\mathchoice
{\hbox{\cmss Z\kern-.4em Z}}{\hbox{\cmss Z\kern-.4em Z}}
{\lower.9pt\hbox{\cmsss Z\kern-.4em Z}}
{\lower1.2pt\hbox{\cmsss Z\kern-.4em Z}}\else{\cmss Z\kern-.4em
Z}\fi}
\def\IB{\relax{\rm I\kern-.18em B}}
\def\IC{{\relax\hbox{$\inbar\kern-.3em{\rm C}$}}}
\def\ID{\relax{\rm I\kern-.18em D}}
\def\IE{\relax{\rm I\kern-.18em E}}
\def\IF{\relax{\rm I\kern-.18em F}}
\def\IG{\relax\hbox{$\inbar\kern-.3em{\rm G}$}}
\def\IGa{\relax\hbox{${\rm I}\kern-.18em\Gamma$}}
\def\IH{\relax{\rm I\kern-.18em H}}
\def\II{\relax{\rm I\kern-.18em I}}
\def\IK{\relax{\rm I\kern-.18em K}}
\def\IN{\relax{\rm I\kern-.18em N}}
\def\IP{\relax{\rm I\kern-.18em P}}

%
\def\inbar{\,\vrule height1.5ex width.4pt depth0pt}

\font\cmss=cmss10 \font\cmsss=cmss10 at 7pt
\def\IR{\relax{\rm I\kern-.18em R}}

%


%

%
\Title{\vbox{\baselineskip12pt\hbox{hep-th/0110268}
\hbox{RUNHETC-2001-31}}}
{\vbox{
\centerline{Deformation Quantization and Quantum Field Theory }
\centerline{on Curved Spaces:}
\centerline{the Case of Two-Sphere}}}
\medskip
\centerline{Chengang Zhou}
\smallskip
\centerline{Department of Physics and Astronomy}
\centerline{Rutgers University }
\centerline{Piscataway, NJ 08855--0849}
\smallskip
\centerline{\tt czhou@physics.rutgers.edu.}
\bigskip
\noindent

We study the scalar quantum field theory on a generic noncommutative two-sphere as a special case of noncommutative curved space, which is described by the deformation quantization algebra obtained from symplectic reduction and parametrized by $H^2(S^2, \QR)$. The fuzzy sphere is included as a special case parametrized by the integer two-cohomology class $H^2(S^2, \QZ)$, which has finite number of degrees of freedom and the field theory has a well defined Hilbert space. When the two-cohomology class is not integer valued, the scalar quantum field theory based on the deformation algebra is not unitary: the signature of the inner product on the space of functions is indefinite. Hence the existence of deformation quantization does not guarantee a physically acceptable deformed geometric background. For the deformation quantization on a general curved space, this obstruction of unitarity can be given by an explicit topological formula. 

\Date{September 2001, second edition}
%
\nref\CDS{A.Connes, M.R.Douglas and A.Schwarz," Noncommutative Geometry and
Matrix Theory:compactification on Tori", JHEP 9802 (1998) 003, hep-th/9711162.}
\nref\SW{N.Seiberg and E.Witten, " String Theory and Noncommutative
 Geometry",JHEP 9909 (1999) 032, hep-th/9908142.}
\nref\KS{A.Konechny and A.Schwarz, ``Introduction to m(trix) theory and noncommutative geometry'', hep-th/0012145.}
\nref\BVF{J.Gracia-Bondia, J.Varilly and H.Figueroa, ``Elements of noncommutative geometry'', Birkhaeuser, Boston, 2001.}
\nref\DN{M.Douglas and N.Nekrasov, ``Noncommutative Field theory'', hep-th/0106048.}
\nref\BFFLS{F. Bayen, M.Flato, C.Fronsdal, A.Lichnerowicz and D.Sternheimer, ``Deformation Theory and Quantization I,II'', Ann. Phys. (NY), (1978) 111, 61-110, 111-151.}
\nref\F{B.V.Fedosov, ``A Simple Geometrical Construction of Deformation Quantization'', J.Diff.Geom. 40 (1994), 213-238; Deformation Quantization and Index Theorem, Mathematical Topics 9, Akademie Verlag, Berlin (1996).}
\nref\K{M. Kontsevitch, ``Deformation Quantization of Poisson Manifolds, I'', q-alg/9709040.}
\nref\CF{A.A. Cattaneo and G.Felder, ``A Path Integral Approach to the Kontsevich Quantization formula'', math-QA/9902090.}
\nref\M{J.Madore, The fuzzy sphere, class. Quant. Grav. 9, 69, 1992.}
\nref\Ra{M.A.Rieffel, ``Questions on Quantization'', quant-ph/9712009.}
\nref\BBEW{M. Bordemann, M.Brischle, C.Emmrich, S. Waldmann,
Phase space reduction for star-products: An explicit construction for $CP^n$, Lett. Math. Phys. 36 (1996) 357-371; q-alg/9503004.}
\nref\BCG{M. Bertelson, M. Cahen and S. Gutt, ``Equivalence of star products'', Class.Quantum Grav. 14(1997) A93-A107.}
\nref\Wald{S. Waldmann, A remark on non-equivalent star products via reduction for $CP^n$, math.qa/9802078.}
\nref\Rb{M.A.Rieffel, ``Proper action of groups on C$^\ast$-algebras, in ``Mappings of Operator Algebras'', H.Araki and R.V.kadison, eds., Birkhauser, Boston Basel Berlin 1990, 141-182.}
\nref\Fa{B.V.Fedosov, Deforamtion quantization and asymptotic operator representation, Funktsional'nyi Analiz i Ego Prilozheniya, 25(1991), 622-637(Russian).}
\nref\DL{M.DeWilde and P.Lecomte, Lett.Math.Phys 7(1983), 487.}
\nref\ARS{A.Y.Alekseev, A. Recknagel, V. Schomerus, Noncommutative world-volume geometries: Branes on $SU(2)$ and fuzzy spheres, hep-th/9908040. }
\nref\GS{V. Guillemin, S. Sternberg, `` Symplectic Techniques in Physics'', Cambridge University Press, 1984. }	
\nref\S{D. Sternheimer, ``Deformation Quantization: Twenty Years After'', math.QA/9809056.}
\nref\GMS{R. Gopakumar, S. Minwalla, A. Strominger, Noncommutative solitons, JHEP 05, 020, hep-th/0003160.}


\newsec{Introduction}

Noncommutative geometry has re-emerged as a focus of intensive research as recent development in string theory reveals the connection of these two subjects\refs{\CDS}\refs{\SW}. The two prime examples, $\QR_{\theta}^n$ and $T^n_{\theta}$, exhibit many new phenomenon of the quantum field theory and string theory on these backgrounds, such as UV/IR mixing, Seiberg-Witten map, new soliton and instanton solutions, etc. (see reviews \refs{\KS}\refs{\BVF}\refs{\DN} for recent developments.) In both cases, the noncommutative geometries are described by deformation quantaization algebras where a constant $B$ field serves as the symplectic two-form needed for the construction. The geometry is not described in the usual way by defining the open sets and distances, but as an object dual to the algebra of functions satisfying certain conditions. This correspondence originates from an extention of Gelfand-Naimark theorem which uniquely relates the compact manifold to the algebra of bounded functions defined on it. The algebra of smooth functions has a deformed noncommutative but still associative multiplication law, which has a string theory interpretation as the open string operator algebra in the large B limit.

It is natural to try to generalize the construction to curved manifolds and repeat the success in the above two flat space cases. The deformation quantization problem \refs{\BFFLS}, which is to find the noncommutative algebra for a generic Poisson manifold, is highly nontrivial and has been developed by mathematicians over the past two decades. There are mainly three approaches. In the first approach\refs{\F}, Fedesov first proved that locally on any symplectic manifold there is always a deformation quantization algebra isomorphic to the standard Moyal algebra. After introducing the symplectic connection on the formal Weyl algebra bundle, the noncommutative algebra on the whole manifold is constructed by essentially patching together local Darboux charts and the corresponding Moyal algebras. A trace formula is obtained explicitly. In the second approach, Kontsevich\refs{\K} deals with the more general case of Poisson manifold and find an explicit multiplication law in the form of a formal series expansion. A path integral formulation of a two dimensional topological sigma model was found later\refs{\CF}. In these two approaches, the nature of formal series expansion makes it difficult to perform explicit calculations, and the analysis of the quantum field theory on these background is hampered, for example the convergence of the infinite sum in the multiplication law is a difficult issue even before any discussion of the quantum field theory problem. Thus the symplectic reduction method, which is the third approach, has an appealing feature of providing closed formulae for the deformed product. Although this construction may not be as general as the previous two approaches, this feature is important for discussing physics problems.

The existence of a meaningful quantum field theory on a generic noncommutative curved manifold, even in the simplest case of two dimensional sphere $S^2$, is still an unsettled problem. On the one hand, the construction of fuzzy sphere\refs{\M} utilizes the representation theory of the $su(2)$ Lie algebra, where only the highest weight representations appear with finite dimensional representation space. The enveloping algebra of the $su(2)$ Lie algebra becomes finite rank matrix algebra. This results in the quantization of the radius of the sphere, $x^2+y^2+z^2=l(l+1), l\in \QZ^+$, which in the semi-classical treatment corresponds to the quantization of the total angular momentum. This background supports a well-defined quantum field theory. But the discrete value of the radius for fuzzy sphere is in sharp contrast to the existence of the deformation quantization algebra for ANY redius of the sphere. Rieffel\refs{\Ra} argued that there is unbounded operators at the generic radius of the noncommutative two-sphere, but in view of the usual appearance of the unbounded operators even in quantum mechanics, this argument does not rule out the possibility of valid quantum field theory convincingly. And the example of the  noncommutative two-torus, where there is a one dimensional continuous moduli space of the noncommutative quantum field theory, seems to suggest otherwise. 

To settle this question, we set out to study the explicit deformation quantization algebra for the generic noncommutative two-sphere and the scalar quantum field theory on it. Mindful of the possiblity that some conditions trivial in the usual commutative geometry can become nontrivial in the noncommutative setup, we try to identify the possible source of obstruction imposed by a sensible quantum field theory. The one dimendional moduli space of the noncommutative algebra on the two-sphere from symplectic reduction\refs{\BBEW} is parametrized by the two-cohomology class $H^2(S^2, \QR)$. Because all the two-forms within the same two-cohomology class determine isomorphic star products\refs{\BCG}, we can choose the $SU(2)$-invariant two forms $B=k\omega$, where $\omega$ is the standard volume form on the sphere and the real constant $k=\int_{S^2}B$ parametrizes the noncommutative algebra. The functions on the sphere are obtained from the homomorphism map of certain functions on noncommutative $\QC^2$ in which $S^2$ is embedded. When $k$ is an integer, this homomorphism serves as a projection operator whose image is the space of the finte rank matrix algebra. This correspnods to the case of fuzzy sphere. At non-integer $k$, no such finte cutoff in the space of functions exists, and all the functions on the sphere are present in the algebra. The Laplacian has a well defined eigenvalue problem, and there is a complete basis of orthogonal functions. The total space of continuous functions decomposes under $su(2)$ action into direct sum of finite dimensional irreduciable representation spaces, but it is not a Hilbert space: it has infinitely many positive and negative norm states and so the inner product is not positive definite. This is the obstruction to a sensible quantum field theory. 

There are many important lessons we can draw from this simple example. First, the physically acceptable noncomutative geometry background is different from the ones merely allowing the existence of the deformation quantization algebra. Although there may well be other criterion, the unitarity requirement alone already severely constraints the possibilities. THe deformation quantization algebra satisfying unitarity condition does not always exist on a general Poisson manifold with arbitrary Poisson structure. Furthermore, we will argue that the unitariry constraint appears as the requirement of asymptotic operator representation in Fedosov's formulism\refs{\Fa}. An immediate result is an important topological obstruction formula. 

The result in this paper also proves the uniqueness of the matrix algebra on the two-sphere. As a physical application, it can be regarded as a D-brane wrapping on the two-sphere. We see the unitarity only allows the matrix degrees of freedom, which is expected from the intepretation of the construction of a $D2$-brane from $N$ $D0$-branes. From this point of view, the theory of a brane on two-sphere is the matrix theory. 

The result of this paper can be generalized to the noncommutative $CP^n$.

The paper is organized as follows. In section 2, after a short review of the symplectic reduction in the classical symplectic geometry, we discuss the generalization of this construction to the level of the deformation quantization algebra. We mainly follow \refs{\BBEW} and explicitly construct the deformation quantization algebra on $CP^n$ at generic radius. In section 3 we start to analyze scalar field theory on the noncommutative sphere. The trace of the algebra is constructed, the Laplacian and the related eigenvalue problem is solved, and the norms of the eigenvectors are explicitly calculated. The unitarity obstruction is found at noninteger radius by explicit calculation. We also discuss its relation to the $su(2)$ representation theory. Section 4 contains the discussions about the lessons we learned from this simple example. After drawing the conclusion that deformation quantization algebra is not enough for a physical theory, we discuss the relation of the unitarity requirement to the strict deformation quantization. In particular, we argue that the unitarity requirement is equivalent to the asymplectic operator expansions, and thus find the topological onstruction formula. Finally we comment on the relation of the construction here to the noncommutative scalar soliton solutions.

\newsec{Deformation quantization algebra on $\QC\QP^n$}

\subsec{Deformation Quantization: An Overview}

Noncommutative geometry originally arises from the study of the operator algebra. According to Gelfand-Naimark theorem, the category of compact topological spaces and the category of commutative unital $C^*$-algebras are isomorphic. Generalizing this correspondence, the noncommutative associative algebra is dual to certain noncommutative space. The ``points'' of the noncommutative space is the spectrum of the algebra. Other topological properties are inferred from the algebra. This greatly enlarges the category of the spaces and makes it possible to test the idea that the spacetime at the Planck scale is fundamentally different from the ordinary space. 

One main source of noncommutative spaces is the deformation quantization\refs{\BFFLS}\refs{\S}. It is the reformulation of quantizating a classical mechanics system. The phase space of classical mechanics is a symplectic manifold, where the symplectic form, by definition a nondegenerate closed two-form, determines a Poisson bracket structure on the smooth functions $C(M)$ on the phase space $M$ by $\{f,g\}= \omega ^{ij}\partial_if\partial_jg$. Deformation quantization seeks a noncommutative product, depending on a formal parameter $h$, in the form $f\ast g=\sum_{n=0}^\infty h^nC_n(f,g)$ for $f,g \in C(M)$, such that $C_0(f,g)=fg, C_1(f,g)-C_1(g,f)=\{f,g\}$, where the $C_n$ are bidifferential operators locally of finite order.

The problem of the existence of the deformation quantization has been studide by methematicians for a long time. In the symplectic case, this is proved by Dewilde and Lecomts\refs{\DL} and Fedosov\refs{\F}. Kontsevich proved its existence for any Poisson manifold\refs{\K}. Thus, the deformation quantization always exists.

By generalized Darboux theorem, locally in any open set of a symplectic manifold, the deformation quantization star product is isomorphic to the Moyal product. This seems to raise a question, for example on the two-sphere, if we take the two-form to be zero at one point, then the deformation quantization is effectively over the sphere minus one point, which is diffeomorphic to the two-plane, and the conclusions of two-plane should be suitable for the two-sphere. Actually, the symplectic condition prevents this situation. On the two-sphere, any two-form $B$ is proportional to the standard volume form $\omega$: $B=f(x)\omega$. Requiring $B$ symplectic implies $f(x)$ can not be zero. Thus the deformation on a curved symplectic manifold is truely a nontrivial task.

There is a gauge equivalence relation among the star produts, which is defined on funtions $C(M)$ as a transformation $T:C(M)\rightarrow C(M), T(f)=f+hg_1^i\partial_if+h^2g_2{ij}\partial_i\partial_jf+...$, such that it is an algebric isomorphism $T(f\ast g)=T(f)\ast'T(g)$. We caution that this equivalence is in the category of formal series in the parameter $h$, and requiring the convergence of the product may have nontrivial result.

There is an isomorphism between the equivalence classes of deformations of Poisson structure on $M$ and the equivalence classes of differentiable deformations of the associative algebra $C^\infty(M)$.

The equivalence classes of the star products is determined by the two-cohomology classes $H^2(M, \QR)$\refs{\BCG}. More precisely, for two star products $\ast$ and $\ast'$ which coincide up to m-th order in $h$, the skewsymmetric part of their difference at (m+1)-th order is determined by a closed 2-form on $M$. If this form is exact, there is a equivalence transformation $T$ such that $\ast$ and $T(\ast')$ coincide up to $m+1$-th order in $h$. In particular, the flat symplectic space $\QR^{2n}$ has only trivial cohomology, and the Moyal products for any constant two form are equivalent. This is the basis for the Seiberg-Witten map. Another example is $CP^n$ whose two-cohomology class is one dimensional, and the star products is parametrized by a formal series in $h$ valued in $H^2(CP^n,\QR)$. Upon taking a specific value of $h$, this is again a particular element of $H^2(CP^n,\QR)$. In particular, we can choose in each two-cohomology class the representative two-form to be proprotional to the standard symplectic form and perform quantization. A question ramains is that, there are possibly many different $H^2(CP^n,\QR)$-valued formal series in $h$, although they have different star product formula, will produce the same element in $H^2(CP^n,\QR)$ upon assigning specific value to $h$. We will see shortly that it is closely related to normalization condition about $D(x,\theta)$ in the equivalence transformation $S$ which will be discussed in the next section. They bring nothing to the equivalence classes of the star product on $\QC\QP^n$. Then in the particular example of $CP^n$, the equivalent classes of star products are paremetrized by $H^2(CP^n,\QR)$.

\subsec{Deformation Quantization Algebra on $\QC\QP^n$ from Symplectic Reduction}

In this paper, we will follow the symplectic reduction approach to discuss the deformation quantization algebra on $\QC\QP^n$ \refs{\BBEW}. In this approach, the sympletic manifold is embedded into a high dimensional (usual flat) symplectic space such that the symplectic form on the curved manifold is the pullback of that on the flat symplectic space. The deformation quantization is constructed via a generalized ``pull-back'' from the star product on the flat space. This will gives an explicit closed form multiplication formula on $CP^n$.

We first recall the geometrical construction of the symplectic reduction. Let $Y$ be a symplectic manifold equipped with symplectic form $\Omega$. At each point $p\in Y$ the tangent space is a symplectic vector space under $\Omega_p$. Assume manifold $X$ is embedded inside $Y$, $i:X\rightarrow Y$, such that the tangent space $TX$ is mapped to the coisotropic subspace of $TY$. (The subspace $W$ of a symplectic vector space $V$ is coisotropic if $W^\bot\subset W$.) Usually this is realized by some constraint equation. The pullback of $\Omega$ to $X$, $\Omega'=i^\ast(\Omega)$, is in general degenerate. To get a submanifold $M$ of $X$ on which the induced two-form is nondegenerate, we define the null subspace of $TX$ as $(TX)_p-(TX)_p^\bot-((TX)_p^\bot)^\bot, \forall p\in M$. It can be proved to be a distribution, and, by Frobenius theorem, it determines a foliation of $X$. If the fibration has constant rank, the quotient space $M$ is a differentiable manifold, and $\Omega'$ descends to a nondegenerate symplectic form on $M$. Thus symplection reduction has two steps, the first is to realize the constraint, and the second step is to perform a quotient operation.

The generalization of the symplectic reduction to the deformation quantization is similiar. Although we will work entirely on the algebra of functions, the geometrical picture of those operations should be clear. 

We start from the flat noncommutative space $\QC^{n+1}$ whose coordinates are $(z_1,...$, $z_{n+1}, \bar{z}_1,...,\bar{z}_{n+1})$, and the constant symplectic form is block diagonal, $\omega={i\over 2}\theta^{-1} dz_i\wedge d\bar{z}_i$. We use the Wick product on the noncommutative $\QC^{n+1}$ which is equivalent to the Moyal produt (here we change the formal parameter $h$ in the usual definition of the star product to $\theta$ which is more familiar to physicists),
\eqn\WickA{
F*G=\sum_{r=0}^\infty {\theta^r \over r!} {\partial^r F \over \partial z_{i_1}...\partial z_i^r}{\partial^r G \over \partial\bar{z}_{i_1}...\partial\bar{z}_i^r}.
}

We define the function $x(z)$ as,
\eqn\Radius{
x(z)\equiv\sum_{i=1}^{n+1} z_i\bar{z}_i.
}
In the following we will vary the radius and keep the $\theta$ constant. The deformation algebra depends on the parameter
\eqn\Per{
k=x\theta^{-1} = \int_{S^2}B,
}
this is the same as changing the periods of the effective B field on $CP^n$ while keep the volume.
 
On any symplectic manifold, there is a correspondence between the Hamiltonian functions and the Hamiltonian vector fields via symplectic 2-form. We say function $j(z, \bar{z})$ generates the flow defined by the vector field $X_j$ if $\omega(j,f)=X_jf=\{j,f\}$ for any function $f$. In particular, the function $j(z, \bar{z}) ={x\over 2}$ generates the $U(1)$ action on $\QC^{n+1}$: $z_i \rightarrow e^{i\alpha} z_i, z_i \rightarrow e^{-i\alpha} \bar{z}_i$, with corresponding vector field $X=x{\partial\over\partial x}=\sum_{i=0}^{n+1}i(z_i{\partial\over\partial z_i}-\bar{z}_i{\partial\over\partial \bar{z}_i})$. Now define a real $2n+1$ dimensional sphere $S^{2n+1}$ by $\{z\in C^{n+1}|x(z)=\mu\}={\rm Ker}(x-\mu)$, which is a $U(1)$ fibration over $CP^n$. Then quotient by $U(1)$ action gives $\QC\QP^n= S^{2n+1}/U(1)$. This is the geometric symplectic reduction, and we will discuss its realization on the algebra of the smooth functions with star product \WickA.

The functions on $\QC\QP^n$ are obtained by the projections of the $U(1)$-invariant functions on $S^{2n+1}$, while functions on $S^{2n+1}$ are obtained from putting constraint $x(z)=\mu$ to the functions on $C^{n+1}$. We define the following two sets of functions.

(1) $N_1\equiv\{f\in \QC^\infty(C^{n+1})|f(\alpha z_i,\bar{\alpha}\bar{z}_i)=f(z_i,\bar{z}_i), \alpha\in \QC\setminus\{ 0\}.\}$. These are functions on $\QC\QP^n$ or equivalently homogeneous functions of order zero on $\QC^{n+1}$. Derict calculate the star product on this set of functions is diccicult, so we turn to $N_2$ which is easy to describe and better behaved under the star product. 

(2) $N_2\equiv\{f\in c^\infty(\QC^{n+1})|[x-\mu, f]=0\}$. These functions are $U(1)$-invariant, and the commutator is generated from the star product. We have $N_1\subset N_2 \subset C^\infty(\QC^{n+1})$. $N_2$ is closed under the Wick product, so is actually a star subalgebra. The generators can be chosen as $\{z_i\bar{z}_j, i,j=1,2,...,n+1\}$.

The whole construction is very similiar to the Dirac quantization with constraint $x=\mu$. Put this constraint on the $U(1)$-invariant functions $N_2$ should do. Obviously there are functions in $N_2$ containing factor $(x-\mu)$ which should be ``set to be zero'' to get the quotient algebra. If the algebra were commutative, we can literally set them to be zero, or equivalently set all the $x$ to be constant $\mu$, to get a valid commutative quotient algebra. This is called quotient of the algebra of $N_2$ by the ideal generated by $x-\mu$. But in the noncommutative algebra, this naive operation will have problem.

To see the problem, we need to study the star product more carefully. Let $f\in N_1$, $F\in N_2$, $R_i(x) \in C^\infty(\QR^+)$(these are called radial functions), and the star product between these functions are as follows,

1) $R \ast F = F\ast R=\sum_{r=0}^\infty {\theta^r \over r!}x^r {\partial^r R(x)\over\partial x^r}{\partial^rF\over\partial x^r}$;

2) $R_1\ast R_2=R_2\ast R_1$;

3) $R\ast f=f\ast R=Rf$.

Under the naive quotient where $x$ is set to be constant $\mu$, the radial function is also set to be a constant. The star product of the radial function and homogeneous functions reduces to the commutative pointwise multiplication. So there wll not be any problem for these multiplication. But the star product between the radial functions is not pointwise. Simply set $x$ to be a number $\mu$ is not consistent, a counterexample is sufficient: $ (x\ast R(x))|_{x=\mu} =\mu R(\mu)+ \theta \mu {dR(x)\over dx}|_{x=\mu} \neq \mu R(\mu)$.

In Dirac quantization, this problem is solved by defining an equivalent relations in $N_2$. This is not satisfactory in the current context of obtaining a quotient algebra, where unambiguous functions and product law are required. The solution is to make an equivalence transformation $S$ on the algebra of functions in one variable $x$, such that the induced nonlocal star multiplication becomes the pointwise multiplication. Then set $x$ to be constant $\mu$ will be legitimate operation. Recall that $x$ is the radius square of $\QC^{n+1}$ and takes value in $\QR^+$, its multiplication law is
\eqn\RStar{
R_1(x)\ast R_2(x)=\sum_{r=0}^\infty \theta^r {x^r\over r!}{\partial^r R_1(x)\over\partial x^r}{\partial^rR_2(x)\over\partial x^r}. 
}
The existence of $S$ is ensured by the general deforamtion theory of the associative algebras. The second local Hochschild cohomology group of the associative algebra of smooth complex-valued functions on $\QR^+$ is isomorphic to $\Gamma(\Lambda^2T\QR^+)$ which is trivial, so up to equivalence relation there is only one multiplication in $\QR^+$. But we will see this transformation will have further remification, as it transforms a local field into a nonlocal field, whose meanings we will make precise in the discussion.

The equivalence transformation $S$ is a formal power series in $x$ and the differential $\partial_x$, $S(x,\partial_x)=\sum_{r=0}^\infty\theta^rS_r(x, \partial_x)$, with the defining property
\eqn\S{
S(R_1\ast R_2)=(SR_1)(SR_2), S_0=1, R_1, R_2: R^\dagger\rightarrow C.
}
Defining the symbol $\hat{S}(x,l)$ by the action on the Fourier mode $e_l(x)=e^{lx}$,
\eqn\Symbol{
\hat{S}(x,l)e_l(x)=(Se_l)(x).
}
The star product between the Fourier modes $e_{l_1}\ast e_{l_2}=e_{l_1+l_2+\theta l_1l_2}$ becomes a functorial equation of the symbols under $S$
\eqn\Sa{
\hat{S}(x,\theta l_1l_2+l_1+l_2)e^{\theta l_1l_2x}=\hat{S}(x,l_1)\hat{S}(x,l_2). 
}
The general solution is\refs{\BBEW}
\eqn\Sb{
\hat{S}(x,l)=exp\{ {x\over\theta}D(x,\theta){\rm log}(1+\theta l)-lx\},
}
where $D(x,\theta)={\rm exp}(\theta C(x,\theta))$ is an arbitrary formal power series in $\theta$ with smooth coefficient functions starting with $1$. In particular, $S(x,\partial_x)x=D(x,\theta)x$. Requiring the radius square $x(z)$ of the sphere be invariant under $S$ fixes $D$ to be trivial, $D(x,\theta)=1$, thus \Sb\ becomes
\eqn\Sd{
\hat{S}(x,l)=exp\{ {x\over\theta}{\rm log}(1+\theta l)-lx\}.
}
The transformation of the monomials in $x$ are
\eqn\Se{\eqalign{
&S(x,\partial_x)x^r =x(x -\theta)(x-2\theta)...(x-(r-1)\theta), r\geq 1; \cr
&S(x,\partial_x)x^{-r}=[(x+\theta)(x+2\theta)...(x+r\theta)]^{-1}, r\geq 1.
}}

We can now tie a loose end to the discussion of the equivalent star products on $\QC\QP^n$ at the end of section 2.1. The star product is parametrized by a formal series in $\theta$ valued in $H^2(CP^n,\QR)$, which corresponds to a particular choice of $D(x,\theta)$ as a formal series in $\theta$. Its effect on the transformations of the monomials in $x$ as seen from \Se, apart from a overal constant factor, is merely a rescaling of the radius $\mu$ to $\mu/D(\mu,\theta)$. Thus there is nothing new to the whole equivalence classes of the star products, it merely change the effective value of the radius. Set $D(x,\theta)$ to be one is a convient choice, and it is enough to give all the equivalence classes of the star products.

In conclusion, the deformation quantization algebra on $CP^n$ at radius square $x=\mu$ is obtained through the following procdures:

1) Take the subalgebra of $U(1)$-invariant functions $N_2$ over $\QC^{n+1}$ with induced star products, and rename it $(\CA, \ast)$. It has linear basis being monomials in $\{x_k\}\equiv\{z_i\bar{z}_j\ ,i,j=1,...,n+1\}$.

2) Transform $(\CA, \ast)$ to an equivalent associative algebra $(S(\CA), \tilde{\ast})$ using $S$. As a differential operator in $x$, $S$ only acts on facotor $x$, while homogeneous functions in $N_1$ are invariant under $S$ and behave like constants.  Any degree $d$ monomial can be written as a homogeneous functions times $x^d$, $\Pi_i(x_i)^{d_i}=x^d\Pi_i({x_i\over x})^{d_i}$, where $\sum_id_i=d$. Thus monomials having the degree transform the same under $S$. Notice that after the transformation, the functions of radius $R(x)$ have pointwise multiplication with each other, and we say the radius direction is ``regularized''. 

3) Finally, take the quotient of $(S(\CA),\ast')$ by the two-sided ideal generated by $x-\mu$. This is simply putting $x$ to be $\mu$ in the multiplication formula. $S(\CA)/(x-\mu)$ is the noncommutative associative algebra on $\QC\QP^n$ at radius square $\mu$.

In particular, the degree $d$ monomials in $x_i$'s transform under $S$ as $S(\Pi_i(x_i)^{d_i})=S(x^d\Pi_i({x_i\over x})^{d_i}))=S(x^d)\Pi_i({x_i\over x})^{d_i}=x(x-\theta)...(x-(d-1)\theta)\Pi_i({x_i\over x})^{d_i}$. After restricted to $x=\mu$, it becomes 
\eqn\SeA{
S(\Pi_i(x_i)^{d_i})=(1-{\theta\over \mu})...(1-{(d-1)\theta\over \mu})\Pi_i(x_i)^{d_i},
}
the only change being an additional factor which depends only on the degree of the monomial. In the following we will combine the map $S$ and the quotient by $x-\mu$ as a single operator $S_\mu$, to project out the noncommutative algebra on two-sphere at radius $x=\mu$. 

\newsec{Noncommutative two-sphere and scalar field theory}

\subsec{Noncommutative algebras on two-sphere}

An algebra is a vector space equipped with a multiplication law compatible with the linear structure. The vector space structure can be specified by a complete basis. A set of generators of the algebra should produce a basis through consecutive multiplication, so it usually contains far less elements than a linear basis.

The noncommutative two-sphere can be embedded in the noncommutative $\QC^2$ with coordinate $(z_1, z_2)$. This is also called Wigner-Jordan construction, originally came from the study of the $su(2)$ representation. The coordinate functions $\{x_+,x_-, x_3\}$ are $U(1)$ invariant functions on $\QC^2$:
\eqn\Embed{\eqalign{
j={x\over 2} = {1\over 2} (\bar{z_1} z_1 + \bar{z_2} z_2),\qquad &x_3= {1 \over 2} (\bar{z_1} z_1 - \bar{z_2} z_2), \cr
x_+ = \bar{z_1} z_2,   \qquad  &x_-= \bar{z_2} z_1. \cr
}}
In the usual commutative algebra, they satisfy the relation
\eqn\ConsA{
x_3^2+x_+x_-=j^2.
}

An obvious linear basis for the space of functions on the two sphere is the set of monomials $\{x_3^mx_+^nx_-^l, l,m,n=1,1,2,...\}$ under the constraint \ConsA . A minimum set of generators is obviously the coordinate functions $\{x_+,x_-, x_3\}$. The whole algebra is determined by the star product between these generators, itself induced from the embedding,
\eqn\DegOne{\eqalign{
x_3 \ast x_3&= x_3^2 + \theta j/2,\cr 
x_3 \ast x_+&= x_3x_+ + \theta x_+/2, \qquad x_+ \ast x_3 = x_3x_+ - \theta x_+/2,\cr
x_3 \ast x_-&= x_3x_+ - \theta x_-/2, \qquad x_- \ast x_3 = x_3x_- + \theta x_-/2,\cr
x_+\ast x_-&= x_+x_- + \theta(j+x_3), \qquad x_-\ast x_+ = x_+x_- + \theta (j-x_3).\cr
}}
They satisfy equation
\eqn\ConsB{
x_3\ast x_3+ {1\over 2}(x_+\ast x_-+x_-\ast x_+)= x_3^2+x_+x_-+{3\over 2}\theta j=j^2+{3\over 2}\theta j, 
}
which under the mapping $S_\mu$ becomes
\eqn\ConsC{
x_3 \tilde{\ast} x_3+ {1\over 2}(x_+\tilde{\ast} x_-+x_- \tilde{\ast} x_+)={\mu\over 2}({\mu\over 2}+\theta).
}
The induced commutators for these generators, $[f,g]=f\tilde{\ast}g-g\tilde{\ast}f$, are
\eqn\sutwo{
[x_3, x_\pm ] = \pm x_\pm , \qquad [x_+, x_- ] = 2 x_3.
}
The $x_i$s are exactly the generators of the $su(2)$ Lie algebra. They acts by ajoint representation on the whole noncommutative algebra.

Notice that we can equally choose another set of vector space basis $\{x_{i(1)}\ast x_{i(2)}\ast ... \ast x_{i(n)} | n=1, 2, ...; i(n)=\{3, +, -\}. \}$. The linear equivalence with the symmetric polynomial basis $\{x_3^lx_+^mx_-^n\}$ is obvious from a series of formulae similiar to \DegOne . We will use the symmetric polynomials to avoid the cubersome ordering in the notation.

Specializing the deformation quantization procedure for $CP^n$ in the last sectin to the two-sphere, we can distinguish two qualitatively distinct classes of quotient algebras depending on the moduli $k=\mu\theta^{-1}$:

(1) When $k={\mu\over\theta}$ is a positive integer, $S_\mu$ is a projection operator, whose image contains polynomials in $x_i$'s up to total degree $k$.
\eqn\Sn{\eqalign{
S_\mu (x^r) &=\theta^r {k! \over (k-r)!}, \qquad {\rm if} \qquad r<k; \cr
&=0, \qquad {\rm if} \qquad r\geq k.
}}
This algebra can be identified with Fuzzy sphere at level $k$. We emphasize that the level $k$ appears naturally as the moduli of the algebra here. This result is exact and no artificial truncation is needed. The quantum field theory on this background has been discussed in many papers.

In the limit $k\rightarrow\infty$, all the polynomials are preserved. Furthermore the multiplication law becomes commutative in the limit. Recall that $k=\int_{S^2} B$, this is the same as large $B$ field limit. Or if we fix the radius of the sphere $\mu$, this is the same as small $\theta$ limit. In view of the similar role of $\theta$ as $\hbar$ in the quantization from equation \WickA, this is the semi-classical limit $\hbar\rightarrow 0$ of the quantum mechanics. 

(2) When $k={\mu\over\theta}$ is noninteger, all the polynomials up to infinite degree are present. As a set, it is the same as the commutative algebra on the two-sphere. but the multiplication law has changed. We have
\eqn\Sna{
S_\mu (x^r)=\theta^r ({\mu\over\theta})({\mu\over\theta}-1)...({\mu\over\theta}-r+1).
}
We will explore the properties of these algebras below.

\subsec{Trace on the noncommutative sphere}

To study quantum field theory on the generic noncommutative two-sphere, we first construct the trace. It is a linear functional over the algebra, satisfying the cyclic property similar to the trace of the matrix algebra: ${\rm Tr}(fgh)={\rm Tr}(hfg)={\rm Tr}(ghf)$ for arbitrary elements $f,g,h$ of the algebra as long as the expression is well defined (i.e. $fgh$ is of trace-class). Further we require it reflect certain symmetry of the underlying geometry, which is $SU(2)$ symmetry in the case of two-sphere.

There is only one natural candidate satisfying above conditions. Under the spherical coordinates $(\theta, \phi)$, it is the integration 
\eqn\SphereIntA{
{\rm Tr}(f)=\int {\rm sin\theta d\theta d\phi} f(\theta,\phi).
}

The map $S_\mu$ only affects the radius $x$ part, while the integration over the sphere only involves the angular part. Therefore these two operations are interchangable. We can employ this property to calculate the trace of a function $\tilde{f}=S_\mu(f)$ as ${\rm Tr}(\tilde{f})=S_\mu({\rm Tr}(f))$. In particular, we have
\eqn\TrProj{
\int {\rm sin\theta d\theta d\phi} S_\mu(f) \tilde{\ast} S_\mu(g) = S_\mu(\int {\rm sin\theta d\theta d\phi} f\ast g), 
}
where $f$ and $g$ are homogeneous functions on $\QC^2$ with the natural star product, while $S_\mu(f)$ and $S_\mu(g)$ are their projections to the noncommutative sphere at $x=\mu$ with induced star product $\tilde{\ast}$.

We choose the linear basis of the space of functions to be $\{x_3^lx_+^mx_-^n\}$, and transform to the spherical coordinates,
$x_3={x\over 2}{\rm cos\theta}, x_\pm= {x\over 2}{\rm sin\theta} e^{\pm i\phi}$,
then seperate the radius and the angular dependence,
\eqn\SphereTransB{
x_3^lx_+^mx_-^n=({x\over 2})^{l+m+n}({\rm cos}\theta)^l({\rm sin}\theta e^{i\phi})^m({\rm sin}\theta e^{-i\phi})^n.	
}

The trace of a monomial is then
\eqn\SphereIntB{
{\rm Tr}[S_\mu(x_3^lx_+^mx_-^n)]= \delta_{m,n}4\pi 2^{-(l+2m+1)}S_\mu(x^{l+2m})B({l+1\over 2}, m+1)
}
for $l$ even, and zero for $l$ odd. $B(x,y)={\Gamma(x)\Gamma(y)\over\Gamma(x+y)}$ is the beta function, and $S_\mu(x^{l+2m})$ is given in \Sn \Sna .

\subsec{Uniqueness of the trace}

We have constructed the trace using the spherical symmetry. An important question is whether there is any other trace on this noncommutative algebra. We will argue that the trace is unique up to a constant function.

This follows from the Fedosov's approach to the deformation quantization problem\refs{\F}. In this approach, a formal Weyl algebra bundle is introduced for any symplectic manifold. This bundle can be regarded as the tangent bundle equipped with the star multiplication in the fiber. Define a abelian connection $D$ on this bundle to be a Weyl algebra-valued connection such that $D^2a=0$ for any section $a$ of the bundle. The kernel of $D$, i.e. $Da=0$, are called the flat sections and denoted by $W_D$. They are closed under the multiplication, and in one-to-one correspondence to the smooth functions over the symplectic manifold. Thus its multiplication law map to the smooth functions gives the deformation quantization algebra, or quantum algebra.

A crucial property of quantum algebra is the generalized form of the Darboux theorem. Classically, Darboux theorem states that on a symplectic manifold there exists local coordinates on a neighborhood of any point such that the symplectic form is given by $\omega=dx^1\wedge dx^2+...+dx^{2n-1}\wedge dx^{2n}$. It can be proved that any quantum algebra $W_D$ is locally isomorphic to a trivial Weyl algebra $W_{D^0}(\QR^{2n})$. Now, on any symplectic manifold, we can take a partition of unity $\{\rho_i(x)\}$ subordinate to a open cover $\{U_i\}$, i.e. $\sum_i\rho_i(x)=1$, $supp \rho_i\in U_i$. Then use the above mentioned bijection between the smooth functions on the manifold and the flat sections on the formal algebra bundle, the $\rho_i$ can be mapped to the flat sections $\hat{\rho}_i$ and form a partition of unity in the quantum algebra $W_D$: $\sum_i \hat{\rho}_i\star a=a$.  Define the trace of $a$ to be $tra=tr\hat{\rho}\star a$. Notice that $\hat{\rho}_i$ as a section of the bundle still has support in $U_i$, so will be $\hat{\rho}\star a$. As $U_i$ is isomorphic to the trivial Weyl algebra whose trace is unique, this is well-defined. The only arbitraryness is in the choice of the partition of unity. It is then easy to verify that two partition of unity give the same trace. 

The deformation quantization algebra from the symplectic reduction is the same as the quantum algebra in Fedosov's approach, thus the trace constructed in the previous section is unique up to a constant density function.

\subsec{Quantization of the scalar fields}

We first write down the action for a scalar field on the noncommutative 2-sphere at radius square $x=\mu$, 
\eqn\Lagrang{\eqalign{
S[\tilde{f}] &= S_\mu({\rm Tr}(K[f]-V[f])), \cr
K[f] &= \partial_tf\ast\partial_tf-[x_3,f]\ast[x_3,f]-[x_+,f]\ast[x_-,f]-[x_-,f]\ast[x_+,f]. \cr
}}
The potential $V[f]$ is a function of $f$ with the multiplication law being Wick product.

We can work out the kinetic part in terms of differential operator as follows. First, using the Wick product formula, and denote the star commutator $[x_i,f]$ as $Ad_{x_i}f$, we find
\eqn\AdjointA{\eqalign{
Ad_{x_3} &= {\theta\over 2}[(\bar{z_1}\partial_{\bar{z_1}}-z_1\partial_{z_1})-(\bar{z_2}\partial_{\bar{z_2}}-z_2\partial_{z_2})], \cr
Ad_{x_+} &= \theta (\bar{z_1}\partial_{\bar{z_2}}-z_2\partial_{z_1}), \cr
Ad_{x_-} &= \theta (\bar{z_2}\partial_{\bar{z_1}}-z_1\partial_{z_2}). \cr
}}
Acting on the functions of $(x_3, x_+,x_-)$, they can be expressed as
\eqn\AdjointB{\eqalign{
Ad_{x_3} &= \theta(x_+\partial_{x_+}-x_-\partial_{x_-}), \cr
Ad_{x_+} &= \theta(2x_3\partial_{x_-}-x_+\partial_{x_3}), \cr
Ad_{x_-} &= \theta(x_-\partial_{x_3}-2x_3\partial_{x_+}). \cr
}}
Transformed into the spherical coordinates, these operators are exactly the differential operator forms of the angular momentum operators,
\eqn\AdjointC{\eqalign{
Ad_{x_3} &= \theta (-i\partial_\phi), \cr
Ad_{x_+} &= \theta (e^{i\phi}\partial_\theta+ie^{i\phi}{\rm ctg}\theta\partial_\phi),\cr
Ad_{x_-} &= \theta (-e^{-i\phi}\partial_\theta+ie^{-i\phi}{\rm ctg}\theta\partial_\phi).
}}

The classical equation of motion for the scalar field $f$ is
\eqn\EofM{
\partial_t^2-[Ad_{x_3}^2(f)+Ad_{x_+}Ad_{x_-}(f)+Ad_{x_-}Ad_{x_+}(f)]=V'[f].
}
Notice that $Ad_{x_i}^2(f)=[x_i,[x_i,f]] \neq Ad_{x_i\ast x_i}$. Next set $V=0$ to find the free propgating modes. The spatial part of the Laplacian is exactly the angular part of the Laplace equation in three dimension, and has a well defined eigenvalue problem. The eigenvectors of the operator $(Ad_{x_3}^2+Ad_{x_+}Ad_{x_-}+Ad_{x_-}Ad_{x_+})$ are called solid harmonics when written in terms of $(x_3, x_+, x_-)$:
\eqn\SolHar{
Y_{lm}=[{2l+1\over4\pi}(l+m)!(l-m)!]^{1\over2} \sum_k{(-x_1-ix_2)^{k+m}(x_1-ix_2)^kx_3^{l-2k-m} \over 2^{2k+m}(k+m)!k!(l-m-2k)!}.
}
$Y_{lm}$ is a degree $l$ homogeneous function in $x_i$, with eigenvalue $l(l+1)\theta$ under the Laplacian $(Ad_{x_3}^2+Ad_{x_+}Ad_{x_-}+Ad_{x_-}Ad_{x_+})$. Under the mapping $S_\mu$, it merely chenges by an overall scaling factor $(1-\theta/\mu)(1-2\theta/\mu)...(1-(l-1)\theta/\mu)$. We will find in the next section that this is crucial in determining the sign of the inner product. Notice that $\{Y_{lm}|l=1,2,...; m=-l,-l+1,...,l.\}$ is also a complete set of linear basis of the space of smooth functions on two-sphere, so the eigenvectors of the Lagrangian expand the whole function space.

\subsec{Inner products}

To have a well defined quantum theory, we need a Hilbert space of quantum states in which the inner product is positive definite, i.e. $(f,f)\geq 0$ for any element $f$ in the space. We define the inner product of two functions $\tilde{f}=S_\mu(f), \tilde{g}=S_\mu(g)$ on the noncommutative two-sphere as
\eqn\InnPro{
(\tilde{f},\tilde{g}) = S_\mu(\int {\rm sin\theta d\theta d\phi} \bar{f}(x)\ast g(x)).
}
Under the complex conjugation, the Wick product behaves as $\overline{f\ast g}=\bar{g}\ast \bar{f}$, and the solid harmonics transforms as
\eqn\Conju{
\bar{Y}_{lm} = (-1)^m Y_{l,-m}.
}

The calculation of the inner product among the solid harmonics will proceed in three steps. First, we prove that $Y_{lm}$'s for different $(l,m)$ are orthogonal to each other. The solid harmonics $\{Y_{lm}\}$ together with the star product can be regarded as tensor operators with the associative multiplication of the quantum operators, and $su(2)$ Lie algebra generators $\{x_i\}$ acts by the star commutators, 
\eqn\Rep{\eqalign{
\sum_i Ad_{x_i}Ad_{x_i}(Y_{lm}) &= \sum_i [x_i,[x_i,Y_{lm}]]_\ast =l(l+1)\theta Y_{lm}, \cr
Ad_{x_3} (Y_{lm}) &= [x_3,Y_{lm}]_\ast = m\theta Y_{lm}.
}}
The star product $Y_{lm} \ast Y_{l'm'}$ can be viewed as coupling of tensor operators and in general produces linear combinations of $Y_{lm}$ satisfying certain selection rules. The trace eaaentially  picks up the $SU(2)$ invariant part, i.e. the $Y_{00}$ part. The coupling of two tensor operators into a rotational invariant is given by Wigner coefficient, $C_{mm'0}^{ll'0}$, thus we have
\eqn\Ortho{\eqalign{
(\tilde{Y}_{lm},\tilde{Y}_{l'm'})&=(-1)^m S_u({\rm Tr}(Y_{l,-m}\ast Y_{l'm'})\cr  &\propto (-1)^m C_{-mm'0}^{ll'0})=\delta_{ll'}\delta_{m,m'}{(-1)^l\over \sqrt{2l+1}},
}}
and the claim is proved.

Next, we prove that all the $Y_{lm}$ for the same $l$ but different $m$ have the same norm. Use the cyclic permutation invariance property of the trace and \Rep , we have
\eqn\NormRecA{
{\rm Tr}([x_+,Y_{lm}]\ast [x_-,Y_{lm}])={\rm Tr}([x_+,Y_{lm}]\ast [x_-,Y_{l,-m}]-2mY_{l,-m}\ast Y_{lm}).
}
By angular momentum relations,
\eqn\AnguRel{\eqalign{
[x_+, Y_{lm}]_\ast = \sqrt{(l-m)(l+m+1)} Y_{l,m+1}, \cr
[x_-, Y_{lm}]_\ast = \sqrt{(l+m)(l-m+1)} Y_{l,m+1}, \cr
}}
and the conjugation of the solid harmonic \Conju , this is equivalent to a recursive relation
\eqn\NormRecB{
(l+m)(l-m+1) (Y_{l,m-1}, Y_{l,m-1}) = (l-m)(l+m+1)(Y_{l,m+1}, Y_{l,m+1})+2m(Y_{lm}, Y_{lm}).
}
Then if $Y_{l,m+1}$ and $Y_{lm}$ have the same norm, $Y_{l,m-1}$ has the same norm. This proves the claim.

Finally, we calculate the norm of $\tilde{Y}_{ll}=S_\mu(Y_{ll})$, where $Y_{ll}=\sqrt{{2l+1\over 4\pi}(2l)!} {x_+^l\over (-2)^l l!}$. We find the following result (see Appendix A):
\eqn\NormRecC{
(\tilde{Y}_{ll}, \tilde{Y}_{ll}) = {(2l+1)!\theta^l\over 2^{2l}} S_\mu(x^l) F(-l,-{\mu\over\theta}-l;l+1;1),
}
where $F(a,b;c;1)={\Gamma(c)\Gamma(c-a-b)\over\Gamma(c-a)\Gamma(c-b)}$ is the hypergeometric function $F(a,b;c;z)$ at $z=1$, and it is positive for the values in \NormRecC . So up to a $l$-dependent positive constant, the sign of the norm square of $\tilde{Y}_{lm}$ is determined by $S_\mu(x^l)=\mu(\mu-\theta)...(\mu-(l-1)\theta)$. Obviously, if $l$ is large enough then it can be negative.

Depending on the radius of the sphere, the inner product behaves qualitatively different for the following two cases:

(1) $k=\mu/\theta \in \QR^+- \QZ^+$:

The norm square of $\tilde{Y}_{lm}$ is positive for all $l<\mu/\theta$, and negetive and positive alternately for all $l>\mu/\theta$. The signature of the inner product is $(+++...++-+-+-+...)$ in the $l$ index. It is not a Hilbert space, thus the noncommutative two-sphere at non-integer ratius is not a valid backgound for the quantum field theory.

(2) $k=\mu/\theta \in \QZ^+$:

All the $\tilde{Y}_{lm}$ strictly satisfy $l<\mu/\theta$, so they all have positive norms. It is a finite dimensional Hilbert space, and the noncommutative two-sphere in this case is a valid background for quantum theory.

The effect on the unitarity comes from the nonlocality caused by the noncommutativity. Before the action of the equivalence transformation $S_\mu$, $x^l$ is always positive for $x>0$ and there is no issue of negative norm. $S_\mu$ transform the $l$ degenerate zeros at $x=0$ of $x^l$ into $l$ different nondegererate zeros at $x=0,\theta,...,(l-1)\theta$. It is exactly this factor $S_\mu(x^l)$ that produces negative norm state for the case (1) which destroies the unitarity.

Certainly, to prove that the noncommutative sphere at non-integer radius is not a valid background for quantum field theory, we need to prove that it does not have a Hilbert module, on which the elements of the algabra act as linear operators, and the star product between those element being the associative multiplication between the operators. Suppose there is such a represtation space with a complete orthogonal basis $\{|n\rangle\}$. Any operator $A$ satisfies ${\rm Tr}(\bar{A}\ast A)=\sum_n \langle n|\bar{A}\ast A|n\rangle =\sum ||A|n\rangle||^2 \geq 0$, so all the negative-normed operators of the algebra should annilate all the states in this Hilbert space. This is equivalent to require a quotient of the algebra (or a subset of it) to keep all the non-negative normed elements while preserving the associative product structure. Simple observations about the multiplication law shows there is no such subalgebra, this proves that noncommutative sphere at non-integer sphere can not sustain any quantum field theory.

A better yet simpler argument that the noncommutative sphere at non-integer radius does not support quantum field theory is via path integral consideration. The kinetic term in the action \Lagrang is the sum of the norm of all the solid harmonics, which will be negative for the negative normed modes. Upon performing the path integral, these modes causes $e^{-S}$ to be unbounded, and the path integral is not well defined. The quantum field theory on such background is thus not well defined.

FInally we discuss the relation to the fuzzy phere from the consideration of $su(2)$ representation theory. The $su(2)$ representation theory states there is no unitary representation for non-integer angular momentum. This argument seems to invalidate the noncommmutative algebra at noninteger radius of the two-sphere without any calculation necessary. But simple check of the results in the previous sections shows that non-integer angular momentum never appears. 

Actually there is no contradiction. There are two seperate issues here. The representation theory only cares about the linear structure of the representaion space, while the algebra has to be closed under multiplication in addition to the linear structure. An algebra can be reducible under the symmetric group while irreducible as an algebra. The is actually ythe case here. $su(2)$ Lie algebra generators act on the space of the functions by adjoint action as $Ad_{x_i}$, not act by regular action. The radius in \ConsA is not the total angular momentum in general. Only when $k={\mu\over\theta}$ is integer and there is a finite cutoff in the space functions, the radius coincides with the maximum angular momtntum. In general, any degree $d$ monomial is transformed under $S$ into polynomials with degree no greater than $d$. We can divide the whole space of functions according to the adjoint action by $x_i$'s: the space of polynomials with highest degree no great than $d$ is an invariant subspace of $su(2)$. When $\mu/\theta =k$, the function space is a direct sum of the representations of $su(2)$ with angular momentum of $l=0,1,2,...,k$. At noninteger radius, the space of the functions is decomposed into the direct sum of the irreducible representations of $su(2)$, with every possible total angular momentum eigenvalue present. Although each seperate eigenspaces corresponding to a particular total angular momentum is a Hilbert space itself, they are not necessarily closed under multiplication. Several of them have to be put together to combined into an algebra by the associative multiplication. In this process, the relative sign of the inner product in these subspaces could be different. This is what happens in case of non-integer $k$. The unitarity obstruction to as associative algebra can not be found by using representation theory of $su(2)$ alone.

\newsec{Discussions}

The deformation quantizaiton algebra on the two-sphere provides a good chance to answer some questions about noncommutative geometry on a curved manifold and quantum field theory. We will also comment on the related issues like its relation to the noncommutative solitons in $C^n$ in this section.

\subsec{Deformation quantization, C*-algebra and strict deforamtion quantization} 

In the origianl commutative case, Gelfand-Naimark theorem established the equivalence between the category of the compact amnifold and the category of comutative unital $C^*$-algebra. By definition, $C^*$-algebra is an involutive Banach algebra for which the follwoing identity holds: $||x*x||=||x||^2$ for any elment $x$. The importance of the $C^*$-algebra requirement is that it can always be represented as bounded linear operators on a Hilbert space. This ensures the existence of an underlying Hilbert space.

On the other hand, in the usual deformation quantization approach to the noncommutative geometry, efforts are mainly put to find noncommutative associative algebra itself obtained from the deformation, whose moduli space is continuous. A priori, these are not necessarily $C^*$-algebras. In particular, we find an explicit example of a deformation quantization algebra, the noncommutative algebra of two-sphere at non-integer radius, which is not a $C^*$-algebra, as it has negative-normed elements and does not have Hilbert module. The unitarity condition,  picks up the integer radius and entails the quantization condition. It seems likely that this quantization is imposed by the compactness of the manifold, as quantum mechanics statess there are only finite degrees of freedom on a compact space.

We should combine the deformation quantization with the requirement of unitarity to build a valid geometric background for quantum field theory. M.A. Rieffel\refs{\Ra} uses the concept of strict deformation quantization. It seeks a noncommutative product defined on a dense subset of $C^\infty(M)$ for a Poisson manifold, and requires a $C^*$-algebra structure. It is stronger than the requirement of the unitarity, as it not only requires the underlying Hilbert module, but the elements in the algebra are represented as bounded linear operators\foot{We thank Greg Moore for discussion about this point.}. Rieffel argued that for the two-sphere with the rotational invariant symplectic structure, there is no strict deformation quantization which preserves the action of $SO(3)$\refs{\Rb}, because of the appearance of the unbounded operators. This certainly rules out $C^*$ algebra structure. (The case of fuzzy sphere is not included as it has only finite number of degrees of freedom, and the functions are not dense in the space of functions on the sphere.) But we know that unbounded operators are common in quantum mechanics, and one would doubt whether requirement of the strict deformation quantization is too stringent and there is still valid quantum theory in this case. Our result in this paper rules out this possibility merely using the unitarity requirement. We do not know if we really need the stronger condition of $C^*$ algebra or just unitarity condition, in addition to the existence of the deformation quantization algebra, to ensure the a valid quantum theory. But unitarity is certainly a necesary conditon, and in the case of two-sphere, it is equivalent to requirement of the $C^*$ algebra structure.

\subsec{Unitarity on general symplectic manifold: AOR}

The study of noncommutative two-sphere shows the importance of the untarity condition. For the general symplectic manifold, it imposes a topological constraint on the symplectic form, as we will see in this section.

We use the Fedosov's Weyl algebra bundle approach. As discussed in setion 3.3, the algebra of deformation quantization comprises the flat sections $W_D(M)$ for a choice of abelian connection $D$. The unitarity requires the existence of a Hilbert module of this algebra, such that the functions of the algebra can be represented as operators on this Hilbert space. This is called asymptotic operator representation, stuided by Fedosov in \Fa.

The technical definition of the asymptic operator representation can be seen in \Fa. Using the index theorem in case of deformation quantization algebra, and in particular by explicit construction, Fedosov proved the sufficient and necessary condition for the existence of the AOR is
\eqn\ObstruC{
\omega-{1\over 2}c_1(T_{\QC}M) \in H^2(M,\QZ).
}

This topological obstruction formula should work for simply connected manifoldonly. The noncommutative torus, having nontrivial $\Pi_1$, avoids this obstruction.

\subsec{Mapping $S$ and the noncommutative soliton on $\QC^n$}

Recenely noncommutative spherical solitons on noncommutative $\QC^n$ are found\refs{\GMS} , which is based on a complete set of Hilbert space projection operators defined as functions over $x=r^2$, where r is the radius. These functions are nilpotents under the star product, $\phi_m(x)\ast\phi_n(x)=\delta_{m,n}\phi_n$. Explicitly, they are $\phi_n (x)=L_n(x)e^{-x/2}, n=0, 1,2,...$, where the $L_n$ is the Legurre polynomial and the $\lambda=1$ normalization is assumed. 

Since these soltions depend only on the radius coordinate with multiplication law under the induced star product, while the deformation quantization algebra on sphere is obtained from an algebraic isomphism between the induced star product on $\QR^+$ and the pointwise multiplication on the same $\QR^+$, it is natural to wonder the connection between these them. We will found they are complementary in a sense.

Consider the image under the mapping $S$. As $S$ is a homomorphism of the associative algebras, we have $S(\phi_m)S(\phi_n)=\delta_{mn}S(\phi_m)$ under the commutative pointwise multiplication. How is it possible? It turns out that $S$ is not well defined over all of $\QC^\infty (\QR^+)$. Precisely those nilpotents $\phi_n(x)$ are mapped to singular fuctions on $\QR^+$, which is infinity when $x/\theta$ is smaller than some positive interger and zero otherwise. So this really creates a set of mutually orthogonal step-like singular functions in the commutative functions over $\QR^+$. Thus the Hilbert space of projection operators constructed from $\phi_n$ is complementary to the noncommutative algebras on $\QC\QP^n$.  

This limitation of algebraic cohomological arguments merely repeats the same theme about the relation between the mathematics and physics. In mathematics it is resonable to abstract some properties and ignore the rest. In physics more restriction may be imposed to get a resonable theory. From the theory of formal algebra deformation, any associative multiplication law on $\QR$ is the same: the equivalence transformation always exists. But in general it is not an algebra isomorphism. The domain and the image of the equivalence transformation can both vary.The cohomological consideration only provides informaiton about the associativity, but in physics we have to consider detailed space of functions allowed in the theory.

\centerline {\bf Acknowledgements}

The author would like to thank Michael Douglas for invaluable discussions and encouragement, without which this work would not have been possible.

\appendix{A}{The norm of $\tilde{Y}_{ll}$}

We calculate the norm of $\tilde{Y}_{ll}=S_\mu(Y_{ll})$ in this appendix. The strategy is to use \TrProj, calculate the product of $\bar{Y}_{ll}$ and $Y_{ll}$ using the star product of $\QC^2$, performing the integration, and finally applying the projection $S_{\mu}$.

First, we have 
\eqn\YLL{\eqalign{
Y_{ll}&=\sqrt{(2l+1)!\over 4\pi} ({-1\over 2})^l {x_+^l\over l!}, \cr
\bar{Y}_{ll}&=\sqrt{(2l+1)!\over 4\pi} ({-1\over 2})^l {x_-^l\over l!}.
}}
The norm square of $\tilde{Y_{ll}}$ is 
\eqn\YLLNorm{
(\tilde{Y}_{ll},\tilde{Y}_{ll}) = S_\mu[\int {\rm sin\theta d\theta d\phi} \bar{Y}_{ll}\ast Y_{ll}].
}
Now we need to calculate the star product of $x_+^l$ and $x_-^l$ via the embedding $x_+=\bar{z}_1z_2$, $x_-=\bar{z}_2z_1$, and use \WickA ,
\eqn\StarPlusMin{\eqalign{
x_-^l\ast x_+^l &= \sum_{k=0}^\infty {\theta^k\over k!} {\partial^k(\bar{z}_2z_1)^l \over\partial \bar{z}_2^k} {\partial^k(\bar{z}_1z_2)^l \over\partial z_2^k}, \cr
&=\sum_{k=0}^l {\theta^k\over k!} ({l!\over(l-k)!})^2 ({x\over 2})^{2l-k} (1+{x_3 \over x/2})^l (1-{x_3 \over x/2})^{l-k}. \cr
}}
Transforming into spherical coordinates and perform the integration, we obtain
\eqn\IntPlusMin{
\int {\rm sin\theta d\theta d\phi} x_-^l\ast x_+^l = 4\pi \sum_{k=0}^l {\theta^k\over k!} ({l!\over(l-k)!})^2 x^{2l-k} {\Gamma(l+1)\Gamma(l-k+1)\over\Gamma(2l-k+2)}.
}
Apply the projection operator $S_\mu$, we have
\eqn\FinalIntA{\eqalign{
(\tilde{Y}_{ll},\tilde{Y}_{ll}) 
&= 2^{-2l}\sum_{l=0}^l \theta^k {(2l+1)!\over(2l-k+1)!}{l!\over(l-k)!k!}S_\mu(x^{2l-k}), \cr
&= 2^{-2l} (2l+1)!\theta^l S_\mu(x^l) \sum_{k=0}^l {l!\over (l-k)!k!} {S_{\mu'}({x\over\theta})^k\over (l+k+1)!}, \cr
}}
where in the second equality we have rearranged the label $l\rightarrow l-k$, and write $S_\mu(x^{l+k}) = S_\mu(x^l) S_{\mu'}(x^k)$, where $\mu'=\mu-l\theta$. Introduce the notation $(a)_k=a(a+1)...(a+(k-1))$, we can express the sum in tems of the hypergeometric function,
\eqn\FinalIntB{\eqalign{
(\tilde{Y}_{ll},\tilde{Y}_{ll}) 
&=  2^{-2l} (2l+1)! \theta^l \theta^l S_\mu(x^l) F(-l,-\mu'/\theta; l+1;1).
}}
Substitute in $\mu'=\mu+l\theta$, we obtain \NormRecC .

$F(a,b;c;z)$ is the hypergeometric function, its general definition being
\eqn\HyperDef{
F(a,b;c;z)=\sum_{k=0}^\infty {(a)_k(b)_k\over (c)_k k!}z^k.
}
When the $a=-l$ and $l$ is a non-negative integer, it is a finite sum
\eqn\HyperDef{
F(-l,b;c;z)=\sum_{k=0}^l {(-l)_k(b)_k\over (c)_k k!}z^k,
}
which is the expression used to obtained \FinalIntB\ from \FinalIntA .

\listrefs
\end